\magnification=\magstep1          
\raggedbottom                     
\tolerance=1000                   
\baselineskip 24 truept           
\parindent 36 truept              
\hsize 6.5 truein                 
\vsize 9 truein                   
\overfullrule=0pt
\def\long#1#2#3#4#5{g'_{#1} = (m^2_{#2} - m^2_{#3}) / #4 \approx #5 {\rm ~GeV}}
\def\u#1{$\underline{\smash{\vphantom{y}\hbox{#1}}}$}

\def\notp{p\kern-4.5pt\hbox{$/$}\ }
\def\bard{d\kern-2.5pt\raise 3pt\hbox{-}}

\hfuzz=10.5pt

\centerline{Dynamical Generation of Linear $\sigma$} 
\centerline{Model SU(3) Lagrangian and Meson Nonet Mixing}
\medskip
\centerline{R. Delbourgo and M. D. Scadron\footnote*{Permanent address:
Physics Department, University of Arizona, Tucson AZ 85721, USA}}
\medskip
\centerline{Physics Department, University of Tasmania, Hobart, Australia 7001}
\bigskip
\noindent
This paper is the SU(3) extension of the dynamically generated SU(2) linear 
$\sigma$ model Lagrangian worked out previously using dimensional
regularization. After discussing the quark-level Goldberger-Treiman relations 
for SU(3) and 
the related gap equations, we dynamically generate the meson cubic and 
quartic couplings.  This also constrains the meson-quark coupling constant 
to $g=2\pi/\sqrt3$ and determines the SU(3) scalar meson masses in a 
Nambu-Jona-Lasinio fashion.  
Finally we dynamically induce the U(3) pseudoscalar 
and scalar mixing angles in a manner compatible with data.
\vskip 20pc
\eject
\noindent
{\bf 1. Introduction}
\medskip
\noindent In a recent paper ref.~[1],
we have extended the original spontaneously broken 
SU(2) linear $\sigma$ model (L$\sigma$M) to the quark-level dynamically 
generated L$\sigma$M.  The latter L$\sigma$M is close in spirit to the 
four-fermion theory of NJL, only the tight-binding bound states with 
chiral-limiting mass $m_\pi=0$ and $m_\sigma=2m_{q}$ in the NJL approach 
become elementary particle states in the L$\sigma$M scheme.  
Dimensional regularization and $Z=0$ compositeness conditions are the key
ingredients making the SU(2)
theory extremely predictive.
In this paper we generalize the dynamically generated L$\sigma$M to 
SU(3) symmetry and also discuss meson nonet mixing.

Experimental signatures [2--4] of the elusive nonstrange isoscalar and strange 
isospinor scalar resonances $\sigma(600-700)$ and $\kappa(800-900)$ 
combined with 
recent theoretical observations on scalar mesons [5,6] make the original SU(2) 
and SU(3) linear sigma model (L$\sigma$M) field theories [7,8] of interest 
once again.  Specifically, a broad nonstrange scalar $\sigma$ (400--900) was extracted in the last reference in [2] and supported in the 1996 PDG tables [3] (with an upper limit mass scale 300 MeV higher).

	Such a scalar $\sigma$ (400--900) has a mean value of $m_\sigma = 650$ MeV,
which is in agreement with the prediction of the dynamically generated L$\sigma$M [1].  This SU(2) L$\sigma$M
 computed in one-loop order reproduces [1,9]
 many satisfying 
chiral-limiting results: $m_\pi=0, \ m_\sigma= 
2m_{q}$ (the latter two of course are true in the four-fermion NJL model 
[10]), vector meson dominance (VMD) universality [11] 
$g_{\rho\pi\pi}=g_\rho$, the dynamically generated scale [1] 
$g_{\rho \pi \pi} = 2 \pi$, the KSRF [12] 
rho mass $m_\rho =\sqrt2g_\rho f_\pi$, and Weinberg's [5] mended chiral 
symmetry decay width relation $\Gamma_\sigma=(9/2)\Gamma_\rho$.  Moreover the 
semileptonic $\pi\to e\nu\gamma$ empirical [3] structure-dependent form 
factors are approximately recovered [13] from the SU(2) L$\sigma$M quark 
and meson loops.  Finally the observed [14] $a_0 (984)\to\gamma\gamma$ 
radiative decay width has been obtained [15] using SU(3) quark and meson 
loops in the L$\sigma$M.

Very recently, the SU(2) L$\sigma$M Lagrangian density 
(shifted around the stable vacuum with $\langle\sigma\rangle=\langle\pi
\rangle=0$ and quarks now with mass $m_q$) having interacting part for
elementary quarks and $\pi$ and $\sigma$ 
mesons,
$${\cal L}_{int}=
g\bar\psi(\sigma+i\gamma_5\vec\tau\cdot\vec\pi)\psi +
g^\prime\sigma(\sigma^2+\vec\pi^2)-\lambda(\sigma^2+\vec
\pi^2)^2/4
\eqno(1a)$$
with chiral-limiting meson-quark and meson-meson couplings [7]
$$g=m_{q}/f_\pi \ , \ \ \ \ \ 
  g^\prime=m^2_{\sigma}/2f_\pi = \lambda f_\pi  
\eqno(1b)$$
(for $f_\pi\approx90$ MeV), has
been {\it dynamically generated} [1].  Such 
dynamical generation is driven by the 
meson-quark interaction $g\bar\psi(\sigma+i\gamma_5\vec\tau\cdot\vec\pi)\psi$ 
alone.  This leads to the chiral-limiting meson masses $m_\pi=0$, $m_\sigma 
=2m_{q},$ meson-meson cubic and quartic couplings $g^\prime$, 
$\lambda=g^\prime/f_\pi$ and also constrains the fundamental 
meson-quark coupling to $g=2\pi/\sqrt{N_c}$.  The latter coupling together
with the NJL
scalar $\sigma$ mass follow from dimensional regularization considerations. However, these results are regularization independent as shown in the second
reference in [1].

Since the analogous (but much more complex) SU(3) L$\sigma$M Lagrangian [8]
has 
only been considered in its (unshifted) spontaneously generated form (but also 
giving rise to interesting physics [16,17]), in this paper we try to 
dynamically generate the SU(3) L$\sigma$M Lagrangian, the U(3) meson masses, 
couplings, and in addition, dynamically induce the empirical meson mixing
angles.

In Section 2 we focus on the quark-level SU(3) Goldberger-Treiman (GT) 
relations and corresponding SU(3) ``gap equations.''  Then in Section 3 we 
dynamically generate the nonstrange (NS) $\sigma$ meson--$\pi\pi$ and 
$K\bar K$ couplings $g^\prime_{\sigma_{NS}\pi\pi}$, 
$g^\prime_{\sigma_{NS}K K}$ obtained from the vanishing 
chiral-limiting pseudoscalar meson masses $m_\pi=0, \ m_K=0$.  The latter 
also gives rise to the strange (S)  $\sigma$ meson--$K\bar K$ coupling 
$g^\prime_{\sigma_S K K}$.  Next in Section 4 we dynamically generate the 
SU(3)-broken scalar meson masses (but with $m_\pi = m_K = 0$)
$$m_{\sigma_{NS}}=2\hat m \ , \ \ \ \ \ 
  m_\kappa=2\sqrt{m_s\hat m} \ , \ \ \ \ \ 
  m_{\sigma_S}=2m_s \ . \eqno(2)$$
Here the nonstrange, kappa and strange scalar meson masses are 
$m_{\sigma_{NS}}, \ m_\kappa, \ m_{\sigma_S}$ and the nonstrange and strange 
constituent quark masses are $\hat m, \ m_s$, respectively.

In Section 5 we comment on ``bootstrapping''  the cubic and quartic 
meson-meson couplings from one-loop order to tree order based on the gap 
equations discussed in Section 2.  Finally, in Section 6 we dynamically 
induce the U(3) quark-annihilation graphs in the SU(3) L$\sigma$M.  They 
simulate (but do not double count) the effects of nonperturbative QCD 
by predicting $\eta-\eta^\prime$ and $\sigma- f_0$ mixing angles 
that in fact are compatible with data.  The latter mixing approach, 
while fitted self-consistently, bypasses a direct nonperturbative calculation 
of the singlet U(3) meson 
masses.  We summarize our dynamically generated findings in Section 7 and 
list in the Appendix the needed nonstrange and strange (quark basis) U(3) 
structure constants.
\bigskip
\leftline{\bf 2. Quark Level GT Relations And Gap Equations}
\medskip
Using only constituent quark masses already induced through vacuum
expectation values of scalar fields along with the meson-quark (chiral)
coupling
$g\bar\psi[(\sigma_{NS}+i\gamma_5\vec\lambda\cdot\vec\pi)$ 
$+~\lambda^a \left(\kappa^a+i\gamma_5K^a\right)$ 
$+\cdots]\psi,$ the quark loop 
pion and kaon decay constants depicted in Figs.~1 are in the  
\noindent chiral limit $(q_\pi\to0, \ 
q_K\to0$ but $m_s\neq\hat m$) with $\bard^4p=d^4p(2\pi)^{-4}$,
$$if_\pi=4N_cg\int{\bard^4p~\hat m\over(p^2-\hat m^2)^2} \ , \ \ \ \ \ 
  if_K=4N_cg\int{\bard^4p{1\over2}(m_s+\hat m)\over(p^2-\hat m^2) 
    (p^2-m^2_s)} \ . \eqno(3a)$$
Then invoking the quark-level 
pion GT relation in (1b) and its natural kaon generalization 
[18],
$$f_\pi g=\hat m \ , \ \ \ \ \ 
  f_Kg={\textstyle{1\over2}}(m_s+\hat m) \ , \eqno(3b)$$
eqs.~(3a) lead to the (log-divergent) CL gap equations
$$1=-i4N_cg^2\int^{\Lambda^2}{\bard^4p\over (p^2-\hat m^2)^2} \ , \eqno(4a)$$
$$1=-i4N_cg^2\int^{\Lambda^{\prime2}}{\bard^4 p\over (p^2-m^2_s)
(p^2-\hat m^2)} 
\ . \eqno(4b)$$

Since in the SU(3) L$\sigma$M, the meson-quark coupling constant $g$ is the 
\u{same} for pions and for kaons (in or away from the CL), the knowledge of 
$g$ in (4a) and in (4b) fixes the log-divergent scales $\Lambda$ and 
$\Lambda^\prime$.  In fact it has been shown [1] in the dynamically 
generated SU(2) L$\sigma$M that for $N_c=3$,
$$g=2\pi/\sqrt{N_c}\approx 3.6276 \ , \eqno(5)$$
compatible with the nonstrange GT estimate in (3b) away from 
the CL $g=\hat m/f_\pi\approx340$ MeV/93 MeV$\approx 3.66$.  (We shall return 
to the derivation of (5) in Section 4.)  Accordingly, the gap equations in 
(4) give the Euclidean integrals 
$$1=\int_0^{\Lambda^2/\hat m^2}{(dq^2/\hat m^2) (q^2/\hat m^2)\over 
    (1+ q^2/\hat m^2)^2} \ \ , \ \ \ \ \ \ \ \ 
  1 \approx \int_0^{\Lambda^{\prime2}/m_s\hat m}{(dq^2/m_s\hat m) 
   (q^2/m_s\hat m)\over (1+q^2/m_s\hat m)^2} \ . \eqno(6)$$
The denominator in the second integral of (6) is the geometric average of the
exact product $(1+q^2/\hat m^2)(1+q^2/m_s^2)$.  Then both integrals
in (6) being unity in turn leads to
$$\Lambda^2/\hat m^2\approx\Lambda^{\prime2}/m_s\hat m\approx 5.3 \ .
\eqno(7)$$

To verify that the dimensionless cutoffs in (7) make physical sense, we must 
first introduce a dimensionful scale.  Returning to the CL nonstrange GT 
relation in (3b), we invoke the CL pion decay constant scale [19] $f_\pi^{CL}
\approx90$ MeV, so that the CL nonstrange quark mass is (with 
divergenceless axial current $\partial 
A^\pi=0$ generating the GT relation for pions)
$$f_\pi^{CL} g=\hat m^{CL}\approx(90 \ {\rm MeV})(3.6276)\approx326 \ {\rm 
MeV} \ . \eqno(8)$$
Away from the CL the ratio of the two GT relations in (3b) is fixed to the 
empirical value [12]
$${f_K\over f_\pi}={1\over2}\left({m_s\over\hat m}+1\right)\approx1.22 \ \ \ \ 
{\rm or} \ \ \ \ {m_s\over \hat m}\approx 1.44 \ . \eqno(9)$$
In fact this constituent quark mass ratio of about 1.4 is also known to hold 
for baryon magnetic dipole moments [20], meson charge radii [21] and 
$K^*\to K\gamma$ decays [22].  However in the chiral limit we might expect 
$m_s/\hat m\leq1.44$, say 4/3.  Finally then the cutoff scales in (7) 
become $$\Lambda\sim\sqrt{5.3}\hat m^{CL}\sim750 {\rm MeV} \ , \ \ \ \ \ \ 
\Lambda^\prime\sim\sqrt{5.3m^{CL}_s\hat m^{CL}}\sim860 {\rm MeV} \ .
\eqno(10)$$

The above 
nonstrange cutoff scale of 750 MeV separates SU(2) L$\sigma$M elementary 
particles (the $u$, $d$ quarks, $\vec\pi$ and $\sigma$ mesons, the
latter taken as $\sigma$ (650) as justified from [2,3] and the discssion in the introduction) from the $\overline{q}q$ bound 
states with mass $>750$ MeV
$(\rho(770),\omega(783),A_1(1260),f_2(1275),A_2(1320))$.  
Likewise the qualitative 
isospinor cutoff scale in (10) of 860 MeV separates elementary $K$(495), 
$\kappa$(820) mesons and $m_s\approx 480$ MeV quarks (as we shall see later) 
from bound state $K^*$(895), $K^{**}$(1350) mesons in the SU(3) L$\sigma$M.
In field theory language, the merging of the elementary particle and bound
state
cutoff scales inferred from the gap equations (4) correspond to $Z=0$ 
compositeness conditions [23], whereby the scalar mesons $\sigma (650)$ and
$\kappa (820)$ 
can be consistently treated either as elementary (in the L$\sigma$M) or as
bound states (in the NJL picture).

\bigskip
\leftline{\bf 3. Dynamical Generation Of The Cubic Meson Couplings}
\medskip
Consider now the Nambu-Goldstone massless pion and kaon in the chiral limit 
(CL).  Starting only with the quark-meson coupling used in Section 2, the 
quark one-loop order pion self-energies are depicted in Figs.~2.  In the CL 
the ``vacuum polarization (VP)''-bubble-type amplitude of Fig.~2a is 
(displaying the two quark propagators in the denominator to parallel
eqs.\ (3) and (4))
$$M^\pi_{VP}=-i4N_c2g^2\int{\bard^4 p(p^2-\hat m^2)\over (p^2-\hat m^2)^2} \ 
, \eqno(11a)$$
where the factor of 2 in (11a) arises from $u$ and $d$ quark loops for 
$\pi^\circ qq$ couplings, or $(\sqrt 2)^2$ for $\pi^+u\bar d$ couplings.  
Likewise the ``quark tadpole'' amplitude of Fig.~2b is in the CL
$$M^\pi_{qktad}={i4N_c2g2g^\prime\over m^2_{\sigma_{NS}}}\int{\bard^4 
p ~\hat m\over p^2-\hat m^2} \ . \eqno(11b)$$
The ``new'' $\sigma\pi\pi$ coupling $2g^\prime$ (the factor 2 is from Bose 
symmetry)
 in Fig.~2b is dynamically generated so that the CL pion mass remains 
zero in one-loop order in the CL:
$$\eqalignno{
     m^2_\pi=M^\pi_{VP}+M^\pi_{qktad}=&0 &(12a) \cr
     \left(g-{2g^\prime\over m^2_{\sigma_{NS}}}\hat m\right)
          \int{\bard^4p\over (p^2-\hat m^2)}=&0 &(12b) \cr
     g^\prime_{\sigma_{NS}\pi\pi} = &m^2_{\sigma_{NS}}/2f_\pi \ . &(12c) \cr}
$$
Note that regardless of the two quadratic divergent integrals in Eqs.~(11), 
the dynamically generated meson-meson tree-level coupling $g^\prime$ in (12c) 
``conspires'' to keep the CL pion massless in (12a) and (12b) [1,9].

This SU(2) L$\sigma$M result (12c) can be 
extended to SU(3) by considering the bubble and quark tadpole graphs in
Figs.~3 
which contribute to the kaon mass.  The bubble amplitude for Fig.~3a is in the 
CL $(p_K\to0, \ m_s\neq\hat m),$
$$M^K_{VP}=-i4N_c2g^2\int{\bard^4p(p^2-m_s\hat m)\over (p^2-m^2_s) 
(p^2-\hat m^2)} \ , \eqno(13a)$$
while the two nonstrange (NS) and strange (S) quark tadpole graphs of Fig.~3b
and Fig.~3c have the respective CL amplitudes 
$$M^K_{qktadNS}=i{4N_cgg^\prime_{NS}\over m^2_{\sigma_{NS}}}\int{\bard^4p2 \hat
m\over (p^2-\hat m^2)} \eqno(13b)$$
$$M^K_{qktadS}=i{4N_c\sqrt2gg^\prime_{S}\over m^2_{\sigma_{S}}}\int{\bard^4p~
m_s\over (p^2-m_s^2)} \ . \eqno(13c)$$

Note that the factor of $2g^2$ in (13a) is due to the two $K^\circ sd$ 
vertices each with coupling $\sqrt 2 g$, while the factor of $2\hat m$ in
(13b) 
counts 2 nonstrange quarks in the NS tadpole loop 
of Fig.~3b.  Finally the factor 
of $\sqrt2g$ in (13c) is due to the $\sigma_{_S} SS$ coupling in 
Fig.~3c.  In order to combine Eqs.~(13a), (13b) and (13c) so that the CL 
kaon mass remains zero,
$$
m^2_K=M^K_{VP}+M^K_{qktadNS}+M^K_{qktadS}=0 \ , 
\eqno(14a)
$$
we invoke the partial-fraction identity
$$
{(m_s+\hat m)(p^2-m_s\hat m)\over (p^2-m^2_s)(p^2-\hat m^2)}={\hat m\over 
p^2-\hat m^2}+{m_s\over p^2-m^2_s} \ . \eqno(14b)
$$
Replacing the integrand of the kaon bubble amplitude in (13a) by the 
right-hand-side (RHS) of (14b), we see that the vanishing of $m^2_K$ in 
(14a) requires the two coefficients of the nonstrange loop integral to 
cancel and also the two coefficients of the strange loop integral to cancel.  
Thus we have dynamically generated {\it two} more tree-level meson cubic
couplings in the 
CL:
$$
g^\prime_{\sigma_{NS}KK}=m^2_{\sigma_{NS}}/2f_K 
\eqno(15a)
$$
$$
g^\prime_{\sigma_{S}KK}=m^2_{\sigma_S}/\sqrt2f_K \ . 
\eqno(15b)
$$

The respective Clebsch-Gordon coefficients of 1, 1/2, 1/$\sqrt2$ in (12c), 
(15a) and (15b) correspond to the SU(3) structure constants 
$d_{NS33}=1$, 
$d_{NSKK}=1/2$, $d_{SKK}=1/\sqrt2$, derived in the Appendix.  Thus the 
dynamically generated cubic meson-meson couplings (12c), (15a) and (15b) 
indeed follow an SU(3) L$\sigma$M pattern.
\bigskip
\leftline{\bf 4.  Scalar Cubic Couplings And Scalar Meson Masses}
\medskip
By chiral symmetry we expect the respective scalar-scalar-scalar meson 
couplings to be identical to the analog scalar-pseudoscalar-pseudoscalar 
couplings.  For the SU(2) case, the two $\gamma_5$ vertices in the 
$\sigma_{NS}\pi\pi$ loop of Fig.~4a reduce the divergence to 
$$
g'_{\sigma_{NS}\pi\pi}=2g\hat m\left[-i4N_cg^2\int{\bard^4p(p^2-\hat m^2)\over 
(p^2-\hat m^2)^3}\right]=2g\hat m \ , 
\eqno(16a)
$$
by virtue of the gap equation (4a).  But the two factors of unity 
replacing the $\gamma_5$'s for the analogue
$\sigma_{NS}\sigma_{NS}\sigma_{NS}$ loop 
of Fig.~4a mean expanding out the trace in the CL gives [1]
$$
  g'_{\sigma_{NS}\sigma_{NS}\sigma_{NS}}  =2g\left[-iN_cg^2 
    \int {\bard^4pTr(\notp+\hat{m})^3 \over (p^2-\hat m^2)^3}\right] 
  ~=~6g \hat{m},  
\eqno(16b)
$$
where we keep only the log divergent piece in the integral of (16b)
since it dominates the coupling constant $g'_{\sigma_{NS}\sigma_{NS}\sigma_{NS}}$.
Only then is the tree-order chiral symmetry $g'_{\sigma_{NS} \pi \pi} =$
$g'_{\sigma_{NS}\sigma_{NS}\sigma_{NS}}$ recovered in one-loop order.

Moreover both cubic loop couplings in (16) ``bootstrap'' the $g^\prime$ tree 
coupling $g^\prime=m^2_{\sigma_{NS}}/2f_\pi$ in the CL provided [1,9] 
$$m_{\sigma_{NS}}=2\hat m, 
\eqno(17a)$$
found by setting $g' = 2g \hat{m} = m_\sigma^2 / 2f_\pi$ and using the GTR
$f_\pi g=\hat m$.  This ``shrinking" of quark loops to points in 
eqs.~(16) again  corresponds to a $Z=0$ 
compositeness condition [23].

	To dynamically generate (17a) we consider 
instead Figs.~5 representing $m^2_{\sigma_{NS}}$ in the CL $p\to0$.
Using dimensional regularization, these graphs sum to [1]

$$
m^2_{\sigma_{NS}} = 16iN_c g^2\int \bard^4 p \Biggl[{ \hat{m}^2 \over (p^2
- \hat{m}^2)^2} -
{ 1 \over p^2 - \hat{m}^2 } \Biggr] = { N_c g^2 \hat{m}^2 \over \pi^2 },
\eqno(17b)
$$
using  a $\Gamma$ function identity $\Gamma (2-l) + \Gamma (1 - l) \to
-1$ in $2l=4$ dimensions. Combining (17a) and (17b) one obtains the meson-quark
coupling $g = 2 \pi / \sqrt{N_c}$.  The latter and (17a) can also be 
dynamically generated via the quark mass gap tadpole combined with the 
above dimensional regularization identity again [1].

In a similar fashion, the scalar kappa meson 
self energies of Figs.~6 have the bubble (VP)
amplitude in the CL
$$M^\kappa_{VP}=-i4N_c(\sqrt2g)^2\int{\bard^4 p(p^2+m_s\hat m)\over 
(p^2-m_s^2)(p^2-\hat m^2)} \ . \eqno(18a)$$
Again adding and subtracting $m_s\hat m$ terms to the numerator of (18a), 
cancelling the quadratic divergent VP part against the tadpole graphs of 
Fig.~6b,c and using the gap equation (4b), we find in the CL
$$m^2_\kappa = 0+2\cdot2m_s\hat m \ \ \ \ \ \hbox{or} \ \ \ \ \ 
m_\kappa = 2\sqrt{m_s\hat m} \ . \eqno(18b)$$
Finally for the purely strange meson self-energy graphs of Figs.~7,  by 
analogy with the
nonstrange scalar mass equation (17b), the graphs sum via the above
dimensionless regularization
identity to 
$$
  m^2_{\sigma_{S}} = 
  8iN_c g^2_S 
  \int \bard^4 p 
  \Bigg[{{m^2_s} \over {(p^2 - m^2_s)^2}} -  
        {{1} \over {p^2 - m^2_s}}
  \Bigg] = 
  {{N_c g^2_S m^2_s} \over {2 \pi^2}}~~.
\eqno(19a)
$$
Invoking the U(3) coupling $g_S = \sqrt{2} g$ (which can be generated via
the strange quark
mass gap), equation (19a) generates the strange scalar meson mass
$$
m_{\sigma_S}=2m_s \ . 
\eqno(19b)
$$
Thus we have dynamically generated the chiral-limiting (NJL-like) 
scalar masses (17a), (18b), (19b) in the SU(3) L$\sigma$M as indicated in 
Eq.~(2).  

There are in fact two independent ways of extending these CL results 
away from the chiral limit so as to obtain the ``physical'' quark and scalar 
meson masses.  More specifically with $m_\pi\neq0$, the nonstrange CL 
relation (16) becomes
$$
m^2_{\sigma_{NS}}-m^2_\pi=(2\hat m^{CL})^2=(653 \ {\rm MeV})^2 
\ \ \ \ \ {\rm or}\ \ \ \ \ m_{\sigma_{NS}}\approx{\rm 668 MeV} \ .
\eqno(20)
$$
Alternatively with $f_\pi\approx 93$ MeV away from the chiral limit, the 
quark-level GT relation (8) becomes (still with $g=2\pi/\sqrt3$),
$$
\hat m=f_\pi g\approx(93 {\rm MeV})(3.6276)\approx337{\rm MeV} \ , 
\eqno(21a)
$$
in close agreement with the $u$ constituent 
quark mass found from magnetic dipole moments 
[20].  Then a NJL-type estimate of the chiral-broken nonstrange scalar
$\sigma$ mass is
$$
m_{\sigma_{NS}}=2\hat m\approx 674  {\rm MeV} \ . 
\eqno(21b)
$$
Henceforth we will take $m_{\sigma_{NS}}\approx670$ MeV as the average 
between (20) and (21b).

The $I=1/2$ scalar kappa meson with mass $m_\kappa\neq0$ follows from an 
``equal-splitting-law'' [24] compared to (20),
$$
m^2_\kappa-m^2_K=m^2_{\sigma_{NS}}-m^2_\pi\approx0.43 {\rm GeV}^2 \ \ \ \ \ 
{\rm or} \ \ \ \ \ m_\kappa\approx820 {\rm MeV} \ . 
\eqno(22)
$$
On the other hand, the chiral-broken strange constituent quark mass found 
from (21a) and the ratio $m_s/\hat m\approx1.4$ from (9) is
$$
m_s=\hat m(m_s/\hat m)\approx 475 {\rm MeV} \ , 
\eqno(23a)
$$
also in reasonable agreement with the magnetic moment determination [20].  
Then the NJL-type estimate of the chiral-broken kappa mass in Eq.~(2) scaled
to (21a) above is
$$
m_\kappa=2\sqrt{m_s\hat m}\approx 2\cdot 337\sqrt{1.4}{\rm MeV}\approx 805~
{\rm MeV} \ , 
\eqno(23b)
$$
with an average $m_\kappa \approx 810$ MeV, midway between (22) and (23b).

In the early 1970's, the particle data group (PDG) suggested the ground state 
kappa mass is in the 800-900 MeV region.  Since 1974, however, this 
$\kappa$ has been replaced by the $\kappa$(1450).  But scaled to the 
$\sigma$(670) mass of (20) or (21), a $\kappa$ in the 800-900 MeV 
region as in (22) and (23) is unavoidable.  Nevertheless it is worth 
commenting on why a (peripherally produced) $\kappa (810)$ has not been
observed.  We suggest it is because of the soft pion theorem [25] (SPT)
suppressing the $A_1 \to \pi (\pi \pi)_{\rm{\bf sw}}$ decay rate due to the
interfering $A_1 \to \sigma \pi$ amplitude.  Likewise, a similar SPT
suppresses the $\kappa (810) \to K \pi$ decay amplitude, explaining why
the PDG tables no longer list the $\kappa (810)$. Specifically, the
latter peripherally produced $\kappa (810)$ in $K^- p \to K^- \pi^+
n$ is suppressed by the quark (box plus triangle) SPT chiral
cancellation as in ref.\ [25].

	Henceforth we will take the ground state kappa mass at $m_\kappa
\approx 810$ MeV as the average between (22) and 
(23b).  It is satisfying that these elusive scalar $\sigma$ and $\kappa$ 
masses were recently seen in polarization measurements(Svec {\it et al.,} 
Refs.~[4]) at 750 MeV and 887 MeV 
respectively, which are unaffected by the above soft pion theorem
of ref.\ [25].

We can also estimate the pure strange $\bar ss$ scalar meson 
mass in two ways.  The equal-splitting-law analogue of (20) and (22), is
$$
m^2_{\sigma_S}-m^2_\kappa=m^2_\kappa-
m^2_{\sigma_{NS}} \ \ \ \ \ {\rm or} \ \ \ \ \ 
m_{\sigma_S}\approx 930{\rm MeV} \ , 
\eqno(24)
$$
while the NJL-like strange scalar mass from (2) using 
$m_s\approx 475$ MeV from (23a) is
$$
m_{\sigma_S}=2 m_s\approx 950 {\rm MeV} \ ,
\eqno(25)
$$
with average mass $m_{\sigma_S} \approx 940$ MeV.  Tornqvist and
Roos in ref.\ [2] claim the $f_0$ (980) is mostly an $\overline{s}s$
scalar meson.  Accounting for the observed scalar mixing angle of
$20^\circ$ (scalar mixing is discussed in eqs.\ (37)--(39) of Section
6), this observed $f_0$ (980) is compatible with the above predicted
$\sigma_S$ (930--950).  The average 
``physical'' chiral-broken scalar meson 
masses which we shall henceforth 
use in our dynamically generated SU(3) L$\sigma$M are then
$$
m_{\sigma_{NS}}\approx670{\rm MeV} \ , \ \ \ \ \ m_\kappa
\approx810{\rm MeV} \ , 
\ \ \ \ \ m_{\sigma_S}\approx 940{\rm MeV} \ . 
\eqno(26)
$$

\eject
\leftline{\bf 5. Bootstrapping the Quartic Meson Lagrangian}

Once the $Z=0$ compositeness 
conditions [23] (or the quark mass gap and meson mass equations) are known, via
the SU(3) gap equations (4) and SU(3) NJL equations (2), to shrink quark
loops to 
L$\sigma$M tree graphs, one should study how to induce the SU(3) quartic 
Lagrangian density
$$
{\cal L}^{L\sigma M}_{quartic} = - \lambda[ \sigma^2_{NS} + \vec{\pi}^2 +
\kappa^2
+ K^2 + \sigma^2_S + \eta^2_S ]^2/4.
\eqno(27)
$$ 

\noindent
The U(2) nonstrange sector of (27) was investigated in ref.[1] via
the $u,d$ quark box of Fig.8a, leading to the chiral limiting (CL) $\pi^o
\pi^o \to \pi^o \pi^o$
amplitude
$$
T = -i 8 N_c g^4 \int \bard^4 p ( p^2 - \hat m^2)^{-2} = 2 g^2,
\eqno(28a)
$$
by virtue of the log-divergent gap equation (4a).  Similarly the 
$\pi^+ \bar{K}^\circ \to \pi^+ \bar{K}^\circ$ quark box of Fig.8b has CL
amplitude
$$
T = 
-i(\sqrt{2})^2 4N_cg^4 \int \bard^4p (p^2-\hat m^2)^{-1} (p^2- m^2_s)^{-1}
[p^2- m_s \hat m] =2g^2, 
\eqno(28b)
$$
by virtue of the partial fraction identity (14b) and gap equation (4).  
Likewise the $\eta_S
\eta_S \to \eta_S \eta_S $ strange quark box of Fig.8c has CL amplitude
$$
T = -i 4 N_c g^4_S \int \bard^4 p ( p^2 -  m^2_s)^{-2} = g^2_S = 2 g^2,
\eqno(28c)
$$
due to the strange quark gap equation analogous to (4a) together with 
$g_S = \sqrt{2} g$.

Thus all three quark box graphs of Figs.8 and equs.(28) have effective
quartic (box) couplings
in the chiral limit
$$
\lambda_{quartic\  box} \to 2 g^2,
\eqno(29a)
$$
whereas the SU(3) L$\sigma$M quartic lagrangian (27) has L$\sigma$M tree
strength
$$
\lambda = g^\prime /f_\pi = (2 \hat m g)/f_\pi = 2 g^2,
\eqno(29b)
$$
by virtue of the quark-level GT relation $\hat m / f_\pi = g$.  The fact
that $\lambda = 2 g^2$
means that, starting from the meson-quark interaction, the SU(3) L$\sigma$M
quark box graphs of
Figs.8 and equs.(28) bootstrap back to the L$\sigma$M quartic Lagrangian
(27).  These are all further
examples of the $Z=0$ compositeness 
condition helping to dynamically generate the entire SU(3) L$\sigma$M
Lagrangian.

\bigskip
\leftline{\bf 6.  Dynamically Inducing Mixing Of Pseudoscalar And Scalar 
Meson States}
\medskip
Thus far, starting from the fundamental SU(3) meson-quark chiral interaction 
$g\bar\psi\lambda^i[S^i+i\gamma_5P^i]\psi$, we have dynamically generated 
the L$\sigma$M cubic and quartic meson-meson couplings, the chiral-limiting 
pseudoscalar and scalar SU(3) meson masses, and even the meson-quark couplings 
$g$ and $g_S$.  Taken together this forms the interacting part of the SU(3)
linear 
sigma model (L$\sigma$M) Lagrangian density
$${\cal L}^{L\sigma M}_{int}={\cal L}_{meson-qk}+
{\cal L}_{meson-meson} \ . \eqno(30)$$

It is then natural to study the additional U(3) mixing Lagrangian 
${\cal L}_{mixing}$.  In the spontaneously generated L$\sigma$M scheme of 
Refs.~[8], such an ``input'' $\cal L$ mixing term introduces extra mixing 
parameters in (30) which are to be 
determined by experiment.  Alternatively in our 
dynamically generated approach to the SU(3) L$\sigma$M, the predicted 
parameters in (30) 
already match observation without introducing new (arbitrary) parameters.  
That is, ${\cal L}_{L\sigma M}$ in (30) is an ``output'' Lagrangian rather 
than an input, and there is no additional ${\cal L}$ mixing Lagrangian.

In this dynamically generated L$\sigma$M theory, the chiral-broken seven 
pseudoscalar (Nambu-Goldstone) meson masses $m^2_\pi$ and $m^2_K$ are 
inserted in the theory by hand and then the six chiral-broken scalar
(NJL-like) masses 
of Eq.~(26) will in turn dynamically generate (fit) the observed $\eta$ and $\eta^\prime$ pseudoscalar along with the $\sigma$ and $f_o$ scalar meson 
masses.

More specifically, for the case of pseudoscalar (P) meson states, the U(3)
meson $\eta-\eta^\prime$ mixing is generated by the quark 
annihilation amplitude $\beta_P$ which turns a $\bar uu$ or $\bar dd$ meson 
$P$ into a $\bar uu, \ \bar dd$ or $ \bar ss$ meson $P^\prime$ state.  This
dynamical 
breaking of the OZI rule can be characterized in the language of QCD 
[20,26], in the model-independent mixing approach of Ref.~[27], or in 
terms of the SU(3) L$\sigma$M.  

To reach the same mixing conclusions (28)-(30) in the context of
QCD, one observes that
a singlet $\eta_o$ is twice formed via the ``pinched" quark annihilation
graph in Fig.8.  But
such quark triangle graphs ``shrink" to points in the L$\sigma$M by the
log-divergent gap equations (4), 
i.e. via $Z=0$ conditions.  Then $I=0$ mesons take the place of QCD gluons in
Fig.8 so that one must consider
the L$\sigma$M meson loop graphs of Fig.9 as 
simulating $\beta_P$ for $\eta^\prime - \eta$ mixing.  In all of the above
cases, one classifies the $I = 0$ nonstrange meson mass matrix as
$$M^2_P=\left(\matrix{m^2_\pi+2\beta_P & \sqrt2\beta_P \cr
                      \sqrt2\beta_P & 2m^2_K-m^2_\pi+\beta_P \cr}\right)
\to\left(\matrix{m^2_\eta & 0 \cr
                 0 & m^2_{\eta^\prime}\cr}\right) \ , \eqno(31)$$

\noindent
where the arrow indicates rediagonalization to the observed $\eta$ and $\eta'~
I = 0$ states.  Here nonstrange (NS) and strange (S) $I = 0$ meson states
contribute to a unitary singlet state according to $| 0 \rangle = \sqrt{2/3} 
| NS \rangle +  {\sqrt {1/3}} | S \rangle$, where $ |NS \rangle = 
|\overline u u + \overline d d \rangle / \sqrt {2}$ and $|S \rangle = |
\overline s s \rangle$.

Note that the \it one \rm parameter $\beta_P$ on the LHS of (31) determines the
\it two \rm measured masses $m^2_\eta$ and $m^2_{\eta^\prime}$ on 
the RHS of (31).  
Specifically, 
the trace of (31) requires
$$2m^2_K+3\beta_P=m^2_\eta+m^2_{\eta^\prime} \ \ \ \ \ {\rm or} \ \ \ \ \ 
  \beta_P\sim0.24 {\rm GeV}^2 \ , \eqno(32a)$$
while the determinant of (31) gives  
$$m^2_\pi(2m^2_K-m^2_\pi)+(4m^2_K-m^2_\pi)\beta_P=m^2_\eta m^2_{\eta^\prime} 
\ \ \ \ \ {\rm or} \ \ \ \ \ \beta_P\sim0.28{\rm GeV}^2 \ . \eqno(32b)$$
Rather than work with isospin L$\sigma$M intermediate states and 
dynamically generate $\beta_P$, it will be more straightforward to consider 
intermediate QCD glue (which is automatically flavor blind and $I=0$) and 
``dynamically fit'' $\beta_P$ in (31) to the scale of (32) even in the 
context of the L$\sigma$M [28].  There is then no need to introduce an
additional SU(3) Lagrangian simulating this mixing; it is already built into
the above quantum-mechanical picture.

Further fine tuning of (32) comes from accounting for the SU(3)-breaking 
ratio of the nonstrange and strange quark propagators in Fig.~9 
via the constituent quark mass ratio $X=\hat m/m_s\approx0.7$ from 
Eq.~(9).  In the latter case, keeping $X$ free and weighting the off-diagonal 
$NS$-$S$ 
$\beta_{P}$ in (31) by $X$ and the $S$-$S \ \beta_P$ by $X^2$, the two  
parameters $\beta_P$ and $X$ on the LHS of this modified 
matrix (31) are uniquely 
determined by the $\eta$ and $\eta^\prime$ masses to be [26]
$$\beta_P={(m^2_{\eta^\prime}-m^2_\pi)(m^2_\eta-m^2_\pi)\over 
4(m^2_K-m^2_\pi)}\approx0.28 \ {\rm GeV}^2 \ , \eqno(33a)$$
$$X\approx0.78 \ . \eqno(33b)$$
These latter two fitted parameters are compatible with (32) and 
with $X\approx 0.7$ from (9).  Thus there is only the one nonperturbative 
parameter $\beta_P\approx0.28$ GeV$^2$ to be explained in our dynamically 
fitted scheme.  In fact this $\beta_P$ can be partially understood from 
a perturbative 2-gluon anomaly-type of graph [29,30].  
Since the SU(3) nonstrange 
and strange pseudoscalar and scalar meson masses have already been 
dynamically generated in (26), our dynamically fitted U(3) extension in (33) 
cannot alter the masses in (26), but instead rotates the states via an 
$\eta^\prime - \eta$ mixing angle.  In effect, $\beta_P = m^2_{\eta_o}/3$
in (33) for
$m_{\eta_o} \approx 915$ MeV regardless of the mixing scheme: QCD in Fig.9
or the L$\sigma$M induced by Figs.~8.

To this end one can recast the nonperturbative fitted 
pseudoscalar mixing scale 
of $\beta_P$ in (33) in terms of the quark nonstrange (NS)-strange (S) 
basis pseudoscalar mixing angle $\phi_P$ with physical $\eta$ and 
$\eta^\prime$ states defined by
$$|\eta\rangle=\cos\phi_P|\eta_{NS}\rangle-\sin\phi_P|\eta_S\rangle \ , 
\ \ \ \ \
|\eta^\prime\rangle=\sin\phi_P|\eta_{NS}\rangle+\cos\phi_P|\eta_S\rangle 
\ , \eqno(34a)$$
or equivalently in terms of the more familiar singlet-octet mixing angle
$$\theta_P=\phi_P-\tan^{-1}\sqrt2 \ . \eqno(34b)$$
Given the dynamically generated structure of the mixing mass matrix (31), 
the pseudoscalar mixing angle in (34) is predicted (fitted) to be [26]
$$\phi_{P}=\tan^{-1}\left[{(m^2_{\eta^\prime}-2m^2_K+m^2_\pi) 
  (m^2_\eta-m^2_\pi)\over(2m^2_K-m^2_\pi-m^2_\eta)(m^2_{\eta^\prime}-
  m^2_\pi)}\right]^{1/2}\approx 42^\circ \ , \eqno(34c)$$
or $\theta_P\approx-13^\circ$ since $\tan^{-1}\sqrt2\approx 
55^\circ$ in (34b).  We believe it significant that the ``world data'' in 
1990 pointed to [27] $\theta_P =-14^\circ\pm2^\circ$ or 
$\phi_P = 41^\circ \pm 2^\circ$, in good agreement with (34c).

In order to reconfirm this $\eta-\eta^\prime$ mixing angle prediction (34c) 
specifically in the context of the L$\sigma$M, we consider the radiative 
decays $\pi^\circ\to\gamma\gamma$, $\eta\to\gamma\gamma$, $\eta^\prime\to
\gamma\gamma$.  In the former case, the usual $u$ and $d$ constituent quark 
triangle graphs lead to the $\pi^\circ\gamma\gamma$ amplitude 
$F\epsilon^{\alpha\beta\gamma\delta}k^\prime_\alpha k_\beta
\epsilon_\gamma^{*\prime} 
\epsilon^*_\delta$ with $F=\alpha/\pi f_\pi$.  This of course is the ABJ 
[30] anomaly amplitude or the Steinberger [31] fermion loop result with 
$g_A=1$ and $N_c=3$ at the quark level.  Moreover this $\pi^\circ\gamma\gamma$ 
amplitude is also the L$\sigma$M prediction since there can be no 
three-pion (loop) correction.  The resulting (L$\sigma$M) decay rate is then 
$$\Gamma(\pi^\circ\gamma\gamma)=m^3_\pi(\alpha/\pi f_\pi)^2/64\pi\approx 
7.6 {\rm eV} \ , \eqno(35a)$$
which is very close to experiment [12] $7.74\pm0.55$ eV.

For $\eta,\eta^\prime\to\gamma\gamma$ decays, however, 
an additional constituent 
strange quark loop must be folded into the $\pi^\circ\gamma\gamma$
(L$\sigma$M) 
rate prediction (35a).  This leads to the decay rate ratios [26,27] 
constrained to recent data in [3,32],
$${\Gamma(\eta\gamma\gamma)\over\Gamma(\pi^\circ\gamma\gamma)}= 
\left({m_\eta\over m_\pi}\right)^3\left({5\over 3}\right)^2\cos^2\phi_P
\left(1-{\sqrt2\over 5}{\hat m\over m_s}\tan\phi_P\right)^2=60\pm 7 \ , 
\eqno(35b)$$
$${\Gamma(\eta^\prime\gamma\gamma)\over\Gamma(\pi^\circ\gamma\gamma)}= 
\left({m_\eta^\prime\over m_\pi}\right)^3\left({5\over 3}\right)^2\sin^2\phi_P
\left(1+{\sqrt2\over 5}{\hat m\over m_s}\cot\phi_P\right)^2=550\pm 68 \ . 
\eqno(35c)$$
Using the constituent quark mass ratio $m_s/\hat m\approx 1.4$ from (9) or 
from Refs.~[21,22], the L$\sigma$M rate ratios in (35b,c) respectively 
predict $\phi_P=45^\circ\pm2^\circ$ and $\phi_P=38^\circ\pm4^\circ$, which 
average to the pseudoscalar $\eta^\prime-\eta$ mixing angle extracted from 
$\eta,\eta^\prime\to\gamma\gamma$ observations, 
$$\phi_P=40.5^\circ\pm3^\circ \ \ \ \ \ {\rm or} \ \ \ \ \ 
  \theta_P=-14^\circ\pm3^\circ \ . \eqno(36)$$
Again we believe it is significant that the phenomenological pseudoscalar 
mixing angle in (36) is compatible with the world average $-14^\circ\pm
2^\circ$ in [27] and with the dynamically fitted value 
$\phi_P\approx42^\circ$ in (34c) obtained from quark-annihilation graphs 
for intermediate QCD states.

As regards the chiral analog scalar mixing angle $\phi_S$, the parallel 
to the $\eta-\eta^\prime$ angle $\phi_P$ in (34a) in the nonstrange-strange 
quark basis is defined via the physical $\sigma-f_o$ states
$$|\sigma\rangle=\cos\phi_S|\sigma_{NS}\rangle-\sin\phi_S|\sigma_S\rangle \ , 
\ \ \ \ \ |f_o\rangle=\sin\phi_S|\sigma_{NS}\rangle+\cos\phi_S|\sigma_S\rangle 
\ . \eqno(37)$$
Instead of the dynamical fitted 
approach to $\phi_S$ obtained through the quark-annihilation 
amplitude $\beta_S$ of Ref.~[26] (yielding $\phi_S\sim17^\circ$), given the 
already determined $\sigma_{NS}$ and $\sigma_S$ L$\sigma$M scalar masses in (26), 
we can find $\phi_S$ via $\langle\sigma|f_0\rangle=0$ and (37):  
$$m^2_{\sigma_{NS}}=m^2_\sigma\cos^2\phi_S+m^2_{f_0}\sin^2\phi_S \eqno(38a)$$
$$m^2_{\sigma_{S}}=m^2_\sigma\sin^2\phi_S+m^2_{f_0}\cos^2\phi_S\ . \eqno(38b)$$
These two equations (38) and the physical mass $m_{f_o}\approx980$ MeV 
constrain $m_\sigma$ and $\phi_S$ to the fitted values (with 
$m_{\sigma_{NS}} \approx 670 {\rm MeV}$, $m_{\sigma_{S}} \approx 940 {\rm
MeV}$)
$$m_\sigma=\left[m^2_{\sigma_{NS}}+m^2_{\sigma_S}-m^2_{f_o}\right]^{1/2}\approx 
610 \ {\rm MeV} \ , \eqno(39a)$$
$$\phi_S=\sin^{-1}\left[{m^2_{f_o}-m^2_{\sigma_{S}}\over m^2_{f_o}-m^2_\sigma} 
\right]^{1/2}\approx 20^\circ \ . \eqno(39b)$$

For the pseudoscalar U(3) nonet $(\pi,K,\eta,\eta^\prime)$, we have used the 
dynamically generated SU(3) L$\sigma$M and have self-consistently computed 
(dynamically fitted) the $\eta-\eta^\prime$ mixing angle 
$\phi_P\approx42^\circ$ via the Nambu-Goldstone $\pi$ and $K$ masses 
leading to the nonperturbative mass matrix (31) and Eqs.~(33) 
or mixing angles in (34).  For the scalar U(3) nonet, however, we started with 
the SU(2) dynamically generated NJL-L$\sigma$M $NS$ and $S$ scalar masses 
(26) and 
used equal-splitting laws to fit the $I=1/2$ kappa mass (squared) half-way 
between as in the averages (26).  Together with the observed $f_o$(980) this 
led to the fitted scalar mixing angle $\phi_S\approx20^\circ$ and a slightly 
mixed $I=0$ $\sigma(610)$ mass in (39).  Note that the nearness of the 
observed $f_o(980)$ to our dynamically generated pure $\bar ss$ scalar mass 
$\sigma_S(940)$ is what is forcing $\phi_S$ in (39b) to be small.  This 
parallels the $\bar qq$ vector case where the (nearby) $\phi(1020)$ vector 
meson is known to be almost purely $\bar ss$ strange.  

All that remains undetermined in the 
latter nonet is the $I=1$ scalar $a_o$ mass.  Again it is the chiral 
equal-splitting laws [ESL] that require [24,33] in analogy with (22),
$$
m^2_{a_o}-m^2_{\eta_{NS}}=m^2_{\sigma_{NS}}-m^2_\pi=m^2_\kappa-m^2_K\approx 
0.43 \ {\rm GeV}^2 \ . 
\eqno(40a)
$$
Then with $\phi_P\approx42^\circ$ so that $m_{\eta_{NS}}\approx760$ MeV by 
analogy with (38), Eq.~(40a) predicts the fitted $I=1~a_o$ mass to be
for $m_{\sigma_{NS}}\approx$670 MeV from (20) and (21), 
$$
m_{a_o}=[m^2_{\sigma_{NS}}-m^2_\pi+m^2_{\eta_{NS}}]^{1/2}\approx 
1.00 \ {\rm GeV} \ . \eqno(40b)$$
Of course this latter predicted mass is presumably the observed [12]
$a_o(984)$.  
The fact that this $I=1$ $a_o(984)$ is near the $I=0$ $f_o(980)$ does 
\u{not} necessarily signal that both the $a_o$ and $f_o$ have the same 
(nonstrange) flavor quarks (as do the $\rho$ and $\omega$ and the $a_o$ 
and $f_o$ in the alternative $\bar qq\bar qq$ scheme [34]).  Rather, our 
$\bar qq$ SU(3) L$\sigma$M picture of a mostly strange $f_o (980)$ and 
nonstrange $a_o(984)$ is based on the (standard) mixing equations of (38) 
and (39) and the (infinite momentum frame) ESLs of Eqs.~(40).  
To support this latter $\bar qq$ L$\sigma$M picture is the known almost 
purely strange vector meson $\phi$(1020) being near this mostly strange 
$f_o$(980).  Moreover, Figs.~13 in the DM2 Collab.~in refs.~[2] also suggests
that the $f_0 (980)$ is composed mostly of $\overline s s$ quarks.

In the context of this same L$\sigma$M, the SU(2) meson-meson 
coupling $g^\prime_{\sigma_{NS}\pi\pi}=(m^2_{\sigma_{NS}}-m^2_\pi)/2g_\pi$ 
can be replaced by 
the SU(3) ESL coupling $g^\prime_{\delta\eta_{NS}\pi}=(m^2_{a_0}
-m^2_{\eta_{NS}}) /2f_\pi$.  Continuing to invoke SU(3) symmetry, one can 
then compute the scalar decay rate ratio as
$$
{\Gamma(f_o(980)\pi\pi)\over\Gamma(a_o(984)\eta\pi)}={3p_{f_o}\over 
2p_{a_o}}\left({\sin\phi_S\over\cos\phi_P}\right)^2={37\pm7 {\rm MeV}\over 
57\pm11{\rm MeV}} \ , \eqno(41a)$$
where the observed rates are taken from the 1992 PDG tables [32].  For 
$\phi_P\approx42^\circ$ from (32c), the above 
(39a) requires (with momentum $p_{f_o}=467$ 
MeV, $p_{a_o}=319$ MeV) 
$$
\phi_S=23^\circ\pm3^\circ \ , 
\eqno(41b)
$$
compatible with (39b).  The 1992 PDG tables [32] takes the 
$a_o \longrightarrow \eta \pi$ rate as $57 \pm 11$ MeV, but the high-statistics $a_o \to \eta
\pi$ rate 
measured by Armstrong {\it et al.,} [35] of $95\pm14$ MeV predicts $\phi_S 
\approx18^\circ\pm3^\circ$ from (41a), more in line with (39b).  

The above ground state $\bar qq$ scalar LSM nonet $(\sigma(610), \kappa(810)$, 
$f_o(980),a_o(984))$ with dynamically fitted mixing angle
$\phi_S\approx20^\circ$ 
is qualitatively different from a $\bar qq\bar qq$ 
four-quark [34] or $\bar KK$ molecule [36] scheme.  However the latter 
objections to a $\bar qq$ picture [37] based on the recent Crystal Ball [14] 
radiative decays $a_o\to\gamma\gamma$ and $f_o\to\gamma\gamma$  can also 
be understood in the SU(3) L$\sigma$M (but not in a pure $\bar qq$ quark 
model).  These narrow scalar decays have $\bar qq$ quark loops which 
interfere {\it destructively} [15] with SU(3) L$\sigma$M meson loops and 
lead to rates $\sim0.5$ keV or smaller as measured [14,32].
\bigskip
\eject
\leftline{\bf 7. Conclusion}
\medskip
In summary, we started in Sec.~2 only with the fundamental SU(3) 
meson-quark  (chiral quark model) interaction 
$${\cal L}_{meson-qk}=
  g\bar\psi\lambda^i
  [S^i+i\gamma_5P^i]\psi \ , \eqno(42)$$
(where $i=0,\ldots8$ and $S_i,$ $P_i$ are scalar and pseudoscalar 
elementary meson fields), with quark level SU(3) Goldberger-Treiman 
relations
$$f_\pi g=\hat m\ , \ \ \ \ \ f_K g={\textstyle{1\over2}}
(m_s+\hat m),  \eqno(3b)$$
ensuring $\partial A^j=0$ for $j=1\ldots8$.  
Then we dynamically generated the log-divergent chiral-limiting gap equations
$$1=-i4N_c g^2\int^\Lambda{\bard^4 p\over(p^2-\hat m^2)^2} \ , \ \ \ \ \ 
1=-i4N_c g^2\int^{\Lambda^\prime}{\bard^4 p\over(p^2-\hat m^2)(p^2-m^2_s)} \ , 
\eqno(4)$$
which self-consistently fixed the cutoffs to
$$\Lambda^2/\hat m^2\approx\Lambda^{\prime2}/m_s\hat m\approx5.3 \ , \eqno(7)$$
so that $\Lambda\sim 750$ MeV, or $\Lambda^\prime\sim 860$ MeV in (10).
This required all $I=0,1$ or $I=1/2$ masses less than 750 MeV, 860 MeV to 
be elementary, such as $\hat m, m_s$ for quarks and $m_\pi, m_K,
m_{\sigma_{NS}}
\sim670$ MeV, $m_\kappa\sim 810$ MeV for pseudoscalar and scalar mesons, 
respectively.  

Next in 
Sec.~3 we dynamically generated the cubic meson SU(3) couplings in the CL, 
$$g^\prime_{\sigma_{NS}\pi\pi}=
  m^2_{\sigma_{NS}}/2f_\pi \ , \ \  
  g^\prime_{\sigma_{NS}KK} = 
  m^2_{\sigma_{NS}}/2f_K \ , \ \ 
  g^\prime_{\sigma_{S}KK} = m^2_{\sigma_{S}}/\sqrt2f_K \ , \eqno(43)$$ 
and analogously for $\sigma\sigma\sigma$-like scalar couplings in Sec.~4
with the additional meson-quark SU(3)-limiting coupling constraint for $N_c=3$
$$g=2\pi/\sqrt{N_c}\approx3.6276 \ . \eqno(5)$$
Recall that the cubic part of the standard [8] spontaneously broken SU(3) 
L$\sigma$M Lagrangian density has the SU(3)-limiting form
$$
  {\cal L}_{cubic}^{\rm L \sigma M}=
  g^\prime d^{ijk} S^i (S^j S^k + P^j P^k)~~.
 \eqno(44a)
$$
On the other hand, our dynamically generated SU(3) L$\sigma$M Lagrangian 
also has the SU(3)-limiting structure of (44a), but away from the SU(3) 
limit it becomes 
$$
  \eqalign{
  {\cal L}_{cubic}^{\rm L\sigma M}&=
   g'_{\pi \sigma_{NS} \pi} \vec{\pi} \cdot \vec{\pi}
	~~\sigma_{NS} + g'_{\pi \delta \eta_{NS}} \vec{\delta} 
         \cdot \vec{\pi} ~~\eta_{NS}
	 + g'_{\pi \kappa K} \bar{K} \vec{\tau}
 	\cdot \vec{\pi}~~ \kappa + g'_{K \delta K} \bar{K} \vec{\tau} \cdot
  	\vec{\delta} ~~K \cr
  &+ g'_{K \sigma_S K} \bar{K} K{\sigma_S} + g'_{K \kappa \eta_S} 
	\bar{K}{\kappa}\eta_S + g'_{K \sigma_{NS} K} \bar{K} 
	K{\sigma_{NS}} +
	g'_{K \kappa \eta_{NS}} \bar{K}{\kappa \eta_{NS}}~~. 
  }
\eqno{(44b)}
$$
Here we use the SU(3) partially-broken L$\sigma$M meson-meson couplings
with massive pseudoscalars [38]
$$
  \long{\pi \sigma_{NS} \pi}
  {\sigma_{NS}} {\pi}
  {2f_\pi} {2.3}~~,
\eqno{(45a)}
$$
$$
  \long{\pi \delta \eta_{NS}}
  {\delta}{\eta_{NS}}
  {2f_\pi}{2.1}~~,
\eqno{(45b)}
$$
$$
  \long{\pi \kappa K}
  {\kappa}{K}
  {2f_\pi}{2.2}~~,
\eqno{(45c)}
$$
$$
  \long{K \delta K}
  {\delta}{K}{2f_K}{3.2}~~,
\eqno{(45d)}
$$
$$
  \long{K\sigma_S K}{\sigma_S}{K}{2 \sqrt{2} f_K}{2.1}~~,
\eqno{(45e)}
$$
and the more severely broken SU(3) L$\sigma$M couplings
$$
  \long{K \kappa \eta_S}{\kappa}{\eta_S}{\sqrt{2} f_K}{0.10}~~,
\eqno{(46a)}
$$
$$
  \long{K \sigma_{NS} K}{\sigma_{NS}} {K} {4 f_K}{0.45}~~,
\eqno{(46b)}
$$
$$
  \long{K\kappa \eta_{NS}}{\kappa}{\eta_{NS}}{4f_K}{0.20}~~.
\eqno{(46c)}
$$

In Section 4 we dynamically generated the NJL-L$\sigma$M chiral-broken 
average scalar meson masses appearing in (44) and (45) as 
$$m_{\sigma_{NS}}=2\hat m\sim670{\rm MeV} \ , \ \ \ \ \ 
  m_\kappa=2\sqrt{\hat mm_s}\sim810{\rm MeV} \ , \ \ \ \ \ 
  m_{\sigma_S}=2m_s\sim940{\rm MeV} \ . \eqno(47)$$
Also in Section 5 the bootstrapping of the SU(3) quartic meson couplings in
the 
Lagrangian were discussed, giving the usual value $\lambda=g^\prime/f_\pi 
\approx 26$ 
in the CL.

Finally in Section 6 we focused on U(3) $\eta-\eta^\prime$ and $\sigma-f_o$ 
particle mixing.  In the original spontaneous broken SU(3) L$\sigma$M [8],  
(undetermined) particle mixing parameters were introduced in the extended 
version of the L$\sigma$M Lagrangian (44a).  However in our dynamically 
generated version of the SU(3) L$\sigma$M, \u{no} additional mixing 
parameters enter the Lagrangian (44b).  Rather, OZI violations generate 
quantum-mechanical particle mixing via the diagonalization of the mass 
matrix (31), but the resulting mixing parameters do not enter the dynamically 
generated Lagrangian (44b).  Instead one dynamically fits the latter 
$\eta-\eta^\prime$ and $\sigma-f_o$ mixing angles, in agreement with 
empirical data.  This gives
$$\phi_P\approx42^\circ \ \ \ \ \ \ \ \ \ \ \phi_S\sim20^\circ \ , \eqno(48)$$
respectively for the U(3) nonets $(\pi(138),K(495),\eta(547), \eta^\prime 
(958))_P$ and $(\sigma(610),\kappa(810),$ $f_o(980),a_o(984))_S$.  There is 
much recent data supporting this above SU(3) L$\sigma$M nonet picture [39].

In short, the theoretical dynamically generated and dynamically 
fitted SU(2) and SU(3) linear sigma model Lagrangians of ref.[14] and here 
appear to give a good description of low energy strong interaction 
physics.    Moreover the above L$\sigma$M picture is a natural generalization 
of the four-quark Nambu-Jona-Lasinio dynamically generated scheme and is 
also compatible with low-energy QCD [40].
\medskip

\noindent {\bf Acknowledgements}
One of the authors (MDS) appreciates discussions with A. Bramon, G. Clement, 
V. Elias,  H. F. Jones, 
and R.~Tarrach and is grateful for partial support from the U. S. Department 
of Energy. This research was also partially supported by the Australian
Research Council.
\eject
\leftline{\bf Appendix}
\medskip
Here we translate standard symmetric SU(3) cartesian structure constants 
$d_{oij}=\sqrt{2/3}\delta_{ij}$, $d_{338}=-d_{888}=1/\sqrt3$, $d_{344}=d_{355} 
=1/2$, $d_{366}=d_{377}=-1/2$, $d_{448}=d_{558}=d_{668}=d_{778}=-1/2\sqrt3$,
to the strange $(S)$-nonstrange (NS) quark basis with
$$|NS\rangle=|{\bar uu+\bar dd\over\sqrt2}\rangle=\sqrt{2\over3}|0\rangle+ 
\sqrt{1\over3}|8\rangle \eqno({\rm A1})$$
$$|S\rangle=|{\bar ss}\rangle=\sqrt{1\over3}|0\rangle- 
\sqrt{2\over3}|8\rangle \ . \eqno({\rm A2})$$
Then one finds 
$$\eqalign{
d_{3NS,S}&=d_{3SS}=d_{33S}=0 \ , \cr
d_{33NS}&=d_{NS,NS,NS}=1 \ , \ \ \ \ \ d_{SSS}=\sqrt2 \ , \cr
d_{0NS,NS}&=d_{0SS}=\sqrt{2\over 3} \ , \ \ \ \ \ 
	d_{8NS,NS}={1\over\sqrt3} \cr
d_{8SS}&=-{2\over\sqrt3} \ , \ \ \ \ \ d_{NSKK}={1\over2} \ , \ \ \ \ \ 
	d_{SKK}={1\over\sqrt2} \ . 
\cr}\eqno({\rm A3})$$
\eject
\leftline{\bf References}
\medskip
\item{1.}
R.~Delbourgo and M.~D.~Scadron, {\it Mod.~Phys.~Lett.~} {\u {A10}}, 251 (1995).
  Here dimensional regularization leads to $g = 2 \pi / \sqrt{3}$ and
  $m_\sigma = 2 m_q$, but in fact these results are regularization 
independent. See R. Delbourgo, M.D. Scadron and A.A. Rawlinson ``Regularizing
the quark-level linear $\sigma$ model''.
\medskip
\item{2.} See, {\it e.g.,} P. Estabrooks, {\it Phys. Rev.} \u{D19}, 2678 
(1979); N. Biswas 
{\it et 
al.,} {\it Phys. Rev. Lett.} \u{47}, 1378 (1981); T. Akesson {\it et al.,} 
{\it Phys. 
Lett.} \u{B133}, 241 (1983); N. Cason {\it et al.,} {\it Phys. Rev.} \u{D28} 
, 1586 (1983); A. Courau {\it et al.,} {\it Nucl. Phys.} \u{B271}, 1 (1986); 
DM2 Collab.~, J.~Augustin et al.~{\it Nucl.~Phys.~}{\u {B320}}, 1 (1989).
M.~Svec, {\it Phys.~Rev.\ } D\u{53}, 2343 (1996); N.~A.~Tornquist
and M.~Roos, {\it Phys.\ Rev.\ Lett.\ } \u{76}, 1575 (1996).
\medskip
\item{3.} Particle Data Group, R.\ M.\ Barnett et al, {\it
Phys.\ Rev.\ } D \u{54}, Part I, 1 (1996).
\item{4.} M. Svec {\it et al.,} {\it Phys. Rev.} \u{D45}, 55, 1518 (1992); D \u{53}, 2343 (1996).
\medskip
\item{5.} S. Weinberg, {\it Phys. Rev. Lett.} \u{65}, 1177 (1990).
\medskip
\item{6.} D. Morgan and M. R. Pennington, {\it Z. Phys.} \u{C48}, 623 (1990); 
{\it Phys.~Lett.~}B{\u {258}}, 444 (1991); {\it Phys.~Rev.~}D{\u {48}},
1185 (1993). 
\medskip
\item{7.} M. Gell-Mann and M. L\'evy, 
{\it Nuovo Cimento} \u{16}, 705 (1960); also see 
V. DeAlfaro, S. Fubini, G. Furlan and C. Rossetti, \u{Currents in Hadron 
Physics} (North-Holland, Amsterdam, 1973) chap. 5; more recently see P.~Ko
and S.~Rudaz, {\it Phys.~Rev.~}D{\u {50}}, 6877 (1994).
\medskip
\item{8.} M. L\'evy, {\it Nuovo Cimento} \u{52A}, 23 (1967); S. Gasiorowicz 
and 
D. Geffen, {\it Rev. Mod. Phys.} \u{41}, 531 (1969).
\medskip
\item{9.} T. Hakioglu and M. D. Scadron, {\it Phys. Rev.} \u{D42}, 941 (1990);
 \u{D43} , 2439
(1991); M. D. Scadron, {\it Mod. Phys. Lett.} \u{A7}, 497 (1992).
\medskip
\item{10.} Y. Nambu and G. Jona-Lasinio, {\it Phys. Rev.} \u{122}, 345 (1961)
 (NJL).
\medskip
\item{11.} J. J. Sakurai, {\it Ann. Phys.} \u{11}, 1 (1960).
\medskip
\item{12.} K. Kawarabayashi and M. Suzuki, {\it Phys. Rev. Lett.} \u{16}, 255
 (1966);
Riazuddin and Fayyazuddin, {\it Phys. Rev.} \u{147}, 1071 (1966) (KSRF).
\medskip
\item{13.} P. Pascual and R. Tarrach, {\it Nucl. Phys.} \u{146}, 509 (1978);
 A. 
Bramon and M. D. Scadron, {\it Europhys. Lett.} \u{19}, 663 (1992).
\medskip
\item{14.} Crystal Ball Collaboration, H. Mariske {\it et al.,} {\it Phys.
Rev.} 
\u{D41}, 3324 (1990).
\medskip
\item{15.} S. Deakin, V. Elias, D. McKeon, M. D. Scadron and A. Bramon,
{\it Mod. Phys. Lett.}, A\u{9}, 995
(1994).
\medskip
\item{16.} G. Clement and J. Stern, {\it Phys. Lett.} \u{B231}, 471 (1989).
\medskip
\item{17.}  J. A. McGovern and M. C. Birse, {\it Nucl. Phys.} \u{A506}, 367 
(1990).
\medskip
\item{18.} S. Gerosimov, S. Journ. {\it Nucl. Phys.} \u{29}, 259 (1979); 
R. Delbourgo 
and 
M. D. Scadron, {\it J. Phys.} \u{G5}, 1621 (1979).
\medskip
\item{19.} M. D. Scadron, {\it Repts. Prog. Phys.} \u{44}, 213 (1981); 
S. A. Coon and 
M. D. Scadron, {\it Phys. Rev.} \u{C23}, 1150 (1981) show that
$1-f^{CL}_\pi / f_\pi = 
m^2_\pi / 8\pi^2 f^2_\pi \approx 0.03$ for $ f_\pi \approx 93$ MeV.
\medskip
\item{20.} A. De R\'ujula, H. Georgi and S. Glashow, {\it Phys. Rev.} 
\u{D12}, 147 (1975);
R. Delbourgo and D. Liu, {\it Phys. Rev.} \u{D53}, 6576 (1996).
\medskip
\item{21.} C. Ayala and A. Bramon, {\it Euro. Phys. Lett.} \u{4}, 777 (1987)
 and 
references therein.
\medskip
\item{22.} A. Bramon and M. D. Scadron, {\it Phys. Rev.} \u{D40}, 3779 (1989).
\medskip
\item{23.} A. Salam, Nuovo Cim. \u{25}, 224 (1962); S. Weinberg, Phys. Rev. 
\u{130}, 776 (1962).
\medskip
\item{24.} M. D. Scadron, {\it Mod. Phys. Lett.} \u{A7}, 669 (1992).
\medskip
\item{25.} A.\ Ivanov, M.\ Nagy and M.\ D.\ Scadron, Phys.\ Lett.\ B\u{273},
  137 (1991).
\medskip
\item{26.} H. F. Jones and M. D. Scadron, {\it Nucl. Phys.} \u{B155}, 409 
(1979); M. D. Scadron, {\it Phys. Rev.} \u{D29}, 2079 (1984).
\medskip
\item{27.} A. Bramon and M. D. Scadron, {\it Phys. Lett.} \u{B234}, 346 
(1990); also 
see N. Isgur, {\it Phys. Rev.} \u{D12}, 3770 (1975).
\medskip
\item{28.} Since the diagonalization of the $2\times2$ mass matrix in (31) 
is really a quantum-mechanical (and not a field-theoretical) procedure, we 
suggest that a ``dynamical fit'' in terms of external L$\sigma$M states and 
internal QCD states is a consistent approach.
\item{29.} R.~Delbourgo and M.~D.~Scadron, {\it Phys.~Rev.~} D{\u {28}},
2345 (1983); S. R. Choudhury and M. D. Scadron, {\it Mod. Phys. Lett.} \u{A1},
535 (1986).
\medskip
\item{30.} S. L. Adler, {\it Phys. Rev.} \u{177}, 2426 (1969); J. S. Bell 
and R. 
Jackiw, {\it Nuovo Cimento} \u{60}, 47 (1969) (ABJ).
\medskip
\item{31.} J. Steinberger, {\it Phys. Rev.} \u{76}, 1180 (1949).
\medskip
\item{32.} Particle Data Group, L.~Montanet {\it et al.}, 
{\it Phys.~Rev.~}D{\u {50}}, Part I, 1173 (1994).
\medskip
\item{33.} M. D. Scadron, {\it Phys. Rev.} \u{D26}, 239 (1982).
\medskip
\item{34.} R. L. Jaffe, {\it Phys. Rev.} \u{D15}, 267, 281 (1977).
\medskip
\item{35.} T. A. Armstrong {\it et al.,} {\it Zeit. Phys.} \u{C52}, 389 (1991).
\medskip
\item{36.} J. Weinstein and N. Isgur, {\it Phys. Rev. Lett.} 
\u{48}, 659 (1982).
\medskip
\item{37.} N. N. Achasov and G. N. Shestakov, {\it Zeit. Phys.} 
\u{C41}, 309 (1988).
\medskip
\item{38.} See {\it e.g.} J. Schechter and Y. Ueda, {\it Phys. Rev.} 
\u{D3}, 2874 (1971).
\medskip
\item{39.} See {\it e.g.} P. Estabrooks, Ref.~[2], N. Cason {\it et al.,}
Ref.~[2]; DM2 Collab.~Ref.~[2]  and M. Svec {\it et al.,} 
Refs.~[4].
\medskip
\item{40.} L.\ Baboukhadia, V.\ Elias and M.\ D.\ Scadron,
  ``Linkage between QCD and the Linear $\sigma$ model'',
submitted for publication (1996).
\eject
\leftline{\bf Figure Captions}
\medskip
\leftline{Fig.~1. Pion (a) and kaon (b) decay constants generated by quark 
loops}
\medskip
\leftline{Fig.~2. Pion bubble (a) and quark tadpole (b) graphs}
\medskip
\leftline{Fig.~3. Kaon bubble (a) and quark tadpole (b,c) graphs}
\medskip
\leftline{Fig.~4. Bootstrap of $g_{\sigma_{NS\pi\pi}}$ quark triangle 
(a) to $g^\prime$ tree (b) graph}
\medskip
\leftline{Fig.~5. Scalar $\sigma_{NS}$ bubble (a) and quark 
tadpole (b) graphs}
\medskip
\leftline{Fig.~6. Scalar kappa bubble (a) and quark tadpole (b,c) 
graphs}
\medskip
\leftline{Fig.~7. Scalar $\sigma_S$ bubble (a) and quark-tadpole 
(b) graphs}
\medskip
\leftline{Fig.~8.  Isoscalar gluon-mediated quark annihilation diagrams 
for intermediate QCD states}
\medskip
\leftline{Fig.~9. Quark box graphs for $\pi^\circ \pi^\circ$ (a), $\pi^+
\bar{\kappa}^o$ (b),
$\eta_s \eta_s$ (c) scattering.}
\end